\documentclass[12pt]{article}
\usepackage{graphics}
\begin{document}
\title{UV-IR coupling in higher derivative gravity}
\author{\normalsize Bob
Holdom\thanks{bob.holdom@utoronto.ca} \\
\small {\em Department of Physics, University of Toronto}\\\small {\em
Toronto, Ontario,}
M5S1A7, CANADA}\date{}\maketitle
\begin{picture}(0,0)(0,0)
\put(310,205){UTPT-01-12}
\put(310,190){hep-th/0111056}
\end{picture}
\begin{abstract}  
We discuss the possible existence of new generic vacuum solutions of Robertson-Walker form in higher derivative gravity
 theories in four dimensions. These solutions
illustrate how a dynamical coupling between very low and very high frequency modes can
occur when the cosmological constant is small.
\end{abstract}
\baselineskip 18pt

The question of whether ultraviolet and infrared physics could in some way be coupled
is intriguing.
It is apparent that nonlinearities in some theory would have to play a
role. But if we take the ultraviolet scale to be the Planck scale then Einstein's
 theory, although nonlinear, is inadequate to approach this question. We should at least consider a derivative expansion of some
underlying theory of gravity, in which the Einstein action is simply the two derivative term. 
We advocate here that a study of \textit{generic solutions} (no fine tuning of initial conditions)
 to \textit{generic actions} (no fine tuning of coefficients in the derivative
expansion) can shed some light on the possibilities, despite the purely classical nature of
 the approach. We find evidence for gravitational solutions
with two widely disparate time scales. The \textit{period} of a long time-scale oscillation is 
inversely related to the \textit{amplitude} of a
Planck time-scale oscillation, and both of these quantities are determined by the size of
the cosmological constant.

We consider a general gravitational action along with a free massless scalar field.\footnote{We
shall see below that higher derivative terms involving the scalar are not important.}
\begin{eqnarray} 
S &=&  \int d^4x\, \sqrt{-g}[
M_{\rm Pl}^2\bigl( -2 \Lambda + R + a\, R^2 \\
& &\qquad
+\;b\, R_{\mu\nu}R^{\mu\nu}  + c\,
R_{\mu\nu\lambda\kappa}R^{\mu\nu\lambda\kappa} + \cdots) - {\textstyle{1\over 2}}
\nabla_\mu\phi\nabla^\mu\phi ] ,
\nonumber\label{aa}
\end{eqnarray}
 We assume that all coefficients of the higher order terms are of order one in units of the
Planck mass $M_{\rm Pl}$; e.g. $a$, $b$, $c$ are of order $M_{\rm Pl}^{-2}$. The only 
small dimensionless parameter is taken to be $|\Lambda|/M_{\rm Pl}^2$.
We consider the spatially flat metric,
\begin{equation} ds^2 = - dt^2 + e^{B(t)} \eta_{ij}\, dx^i dx^j ,
\label{dd}\end{equation}
along with $\phi=\phi(t)$. The two gravitational field equations
together imply the scalar field equation, ${\ddot\phi}+\textstyle{3\over 2}\dot B\dot\phi=0$.

An equation for $B(t)$ alone can be obtained. Each
term in this equation, besides the 
cosmological constant term, will have an even number of derivatives acting on some number of
$B(t)$'s, where each $B(t)$ has at least one derivative.
The nonlinearity in the equation is essential for the solutions we study, 
but for $|\Lambda|/M_{\rm Pl}^2\ll
1$ it will turn out that all derivatives of $B(t)$ are of order
$\sqrt{|\Lambda|}$ times the appropriate power of $M_{\rm Pl}$. Thus our solutions
are hardly affected if we drop terms higher order than quadratic in $B(t)$ in the equations.

In addition we shall begin by dropping terms higher order than $R^2$ in (\ref{aa}). This truncation is not
in any way justified, and in fact a problem will arise from this truncation.
But the truncated theory serves to illustrate the basic dynamics, and we shall later 
see how the effects of higher order terms can eliminate the problem.
The truncated equation can be put in the following dimensionless form involving $A(\xi)=B(t)$,
\begin{equation}A''''+A''+{c}_{1}A'^{2}-{c}_{2}A'A'''+{c}_{3}A''^{2}  =  \eta\,\varepsilon^2
,\label{bb}\end{equation} 
where $\xi=t/\sqrt{6a+2b+2c}$, 
$\varepsilon^2=4(3a+b+c)|\Lambda|$, and $\eta={\rm sign}(\Lambda)$. In addition the gravity-scalar action
fixes the nonlinear terms, 
\begin{equation}c_1=3/2\:\:\:,\:\:\:c_2=-9/2\:\:\:,\:\:\:c_3=3/2.\label{ff}\end{equation}
We have assumed that $3a+b+c>0$; this is the only combination of $R^2$ coefficients that
survives for a conformally flat metric, and we have absorbed it into the definition of our
dimensionless expansion parameter $\varepsilon$.
The massless scalar contribution to the energy-momentum tensor corresponds to 
an equation of state $p=w\rho$ with $w=1$;
for a more general contribution we find 
\begin{equation}c_1=3(1+w)/4\:\:\:,\:\:\:c_2=-3-3w/2\:\:\:,\:\:\:c_3=9/4-3w/4.\label{hh}\end{equation}
We shall examine solutions of (\ref{bb}) and treat (\ref{ff}) and (\ref{hh}) as special cases.

The de Sitter expansion $A(\xi)\propto \xi$ is clearly a solution to (\ref{bb})
when ${\rm sign}(c_1)={\rm sign}(\Lambda)$. It can
 also be a generic solution in the sense that for a range of initial conditions the solution will
tend towards the exponential expansion asymptotically. By a numerical analysis this
 is found to occur when
\begin{equation}-{\rm sign}(2{c}_{1}+{c}_{2})= {\rm sign}(c_1)={\rm sign}
(\Lambda).\label{gg}
\end{equation}
The gravity-scalar action satisfies these conditions when $\Lambda>0$,
as does (\ref{hh}) with $w>-1$. The initial conditions
may be such as to produce fast oscillations initially, but these oscillations gradually
die away as illustrated by the top curve in Fig.~(1A).\footnote{The unphysical case
in (\ref{gg}) with $\Lambda<0$ would have exponential \textit{contraction} as the generic solution.}

Fig.~(1) also illustrates a very different set of generic nonsingular solutions that exist when
\begin{equation}{\rm sign}(2{c}_{1}+{c}_{2})= {\rm sign}(c_1+{c}_{2}+{c}_{3})={\rm sign}
(\Lambda).
\end{equation}
By generic we again mean that if $A'(0)$, $A''(0)$, and $A'''(0)$ are each in a respective
 range of values, all of order
$\varepsilon$, then one of these solutions results.\footnote{Derivatives
of $A(\xi)$ being of order $\varepsilon$ is related to the constants $c_i$ being of order one.} We see that the gravity-scalar action
satisfies these conditions as well, this time when $\Lambda<0$.
The same is true for a more general equation of state as long as $w>0$.

These new solutions are characterized by
two scales of periodicity. One is an oscillation with a frequency of order unity in Planck units,
 as determined by the linear terms in (\ref{bb}), and an amplitude
of order $\varepsilon$. The other is a long time-scale oscillation with a frequency of order
$\varepsilon$ and an amplitude of order unity. 
The amplitude of the fast oscillation is modulated with this same period.
The complete solution is not strictly periodic since the phase of the
fast oscillation in general changes after one period of the slow oscillation.
\begin{figure}
\centering
\includegraphics{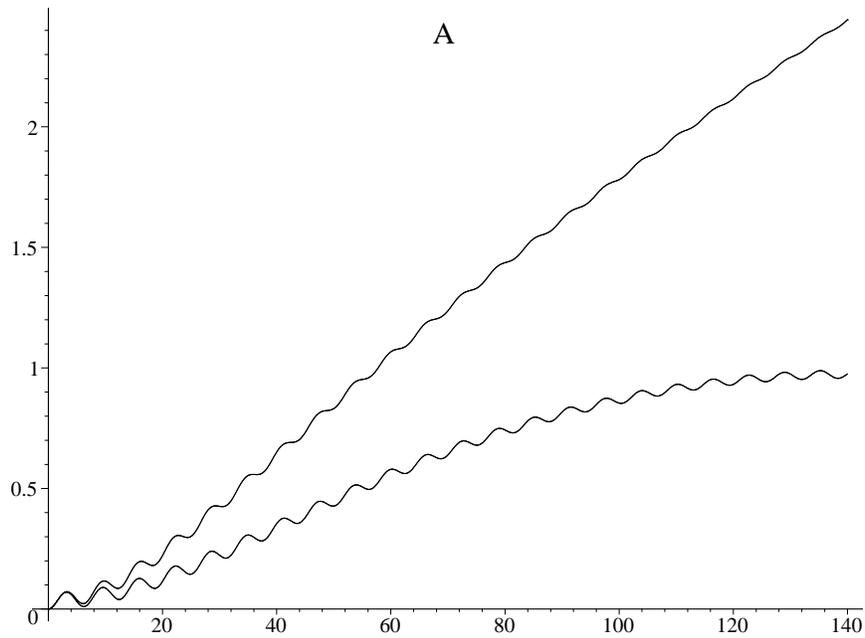}\vspace{10mm}
\includegraphics{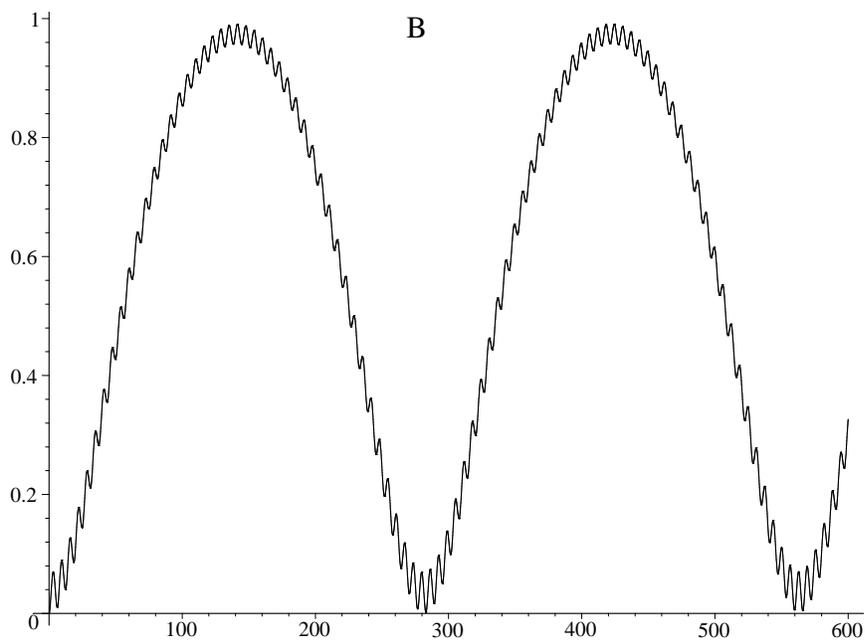}
\caption{Solutions to equation
(\ref{bb}) with the coefficients in (\ref{ff}). In Figure A the top and bottom curves correspond to
$\eta\varepsilon^2=10^{-3}/3$ and $-10^{-3}/3$ respectively. Figure B shows the
continuation of the bottom curve in Figure A. Initial conditions have derivatives vanishing
except for $A''(0)=0.035$. Solutions with properties similar to these occur
for a range of the initial conditions.}
\end{figure}

The two scales of periodicity can be separated arbitrarily far since $\varepsilon$ can
be arbitrarily small. For example if $\Lambda$ is a major component driving the present
evolution of the universe, then the period of the slow oscillation can be of order
the age of the universe. Given that the solution also requires a  fast oscillation of order the Planck time,
it is appropriate to refer to these solutions as UV-IR coupled.

Two observations: 1) For a fine tuning of the initial conditions the amplitude of the
 slow oscillation can be
brought to zero, leaving the fast oscillations only; these are the solutions studied
in section I of \cite{c}. 2) When $\varepsilon$ is not very small then the frequency of the
two oscillations become similar, and some tuning of the initial conditions
becomes necessary. 

Thus far we have considered an equation with up to 4-derivatives only.
As we have alluded to above there is a
problem with the UV-IR coupled solutions in this truncated theory.
The other gravitational field equation determines an oscillating $\dot\phi^2$ that automatically 
satisfies the scalar
field equation, but it also implies that
$\dot\phi^2$ is negative definite.\footnote{In the de Sitter expansion example of Fig.~(1A) 
the $\dot\phi^2$ is also negative while asymptotically vanishing, but there is a 
different range of initial conditions that can
produce positive $\dot\phi^2$.
A change in initial conditions cannot produce positive $\dot\phi^2$ for the UV-IR coupled 
solutions to (\ref{bb}).}  This result is connected
 with having $\Lambda<0$, since the contributions to the pressure $p$ from $\Lambda$
and the average $\dot\phi^2/4$ tend to cancel. Since $\dot\phi^2$ contributes to $\rho+p$ 
(unlike $\Lambda$) this problem with the truncated theory could be avoided if 
there was some other negative contribution 
to $\rho+p$. Higher order derivative terms involving the scalar field cannot
help in this way since
 derivatives of $\dot\phi^2$ are of order
$\varepsilon\dot\phi^2$ in Planck units (as can be seen from the scalar field equation), and $\dot\phi^2$ itself is of order $\varepsilon^2$.

On the other hand all derivatives of $A(\xi)$ are of order $\varepsilon$ and thus
 contributions from higher powers of $R$
can completely change the picture. They will add new linear and nonlinear terms to (\ref{bb})
of the same order in $\varepsilon$, respectively, as the ones in (\ref{bb}).
We have studied the following equation with six derivative terms, where in general
 there is one additional
linear term and three more quadratic terms. 
\begin{eqnarray}d_1 A''''''+A''''+A''+{3\over 2}A'^{2}+{9\over 2}A'A'''+
{3\over 2}A''^{2}&&\nonumber\\+\;d_2 A'''^2-d_3A''A''''+d_4 A'A'''''  &=&  \eta\,\varepsilon^2
\label{ee}\end{eqnarray}
For the two and four derivative terms we have inserted the values in (\ref{ff}).
We find that for a range of
coefficients of the new terms that UV-IR coupled solutions are once again generic,
even when $\Lambda>0$.

The linear terms must support the fast oscillations; that is, the equation for the frequency
obtained by inserting a sinusoid into the linear terms must have real roots only. 
Here it implies that $0<d_1<1/4$. Then to ensure that
 $\Lambda$ and $\dot\phi^2$ are positive, the quadratic
coefficients must be such that $d_2+d_3$ is sufficiently large and positive and $d_4$ is in
some finite range that includes zero.
It seems likely that these conditions can be satisfied for some range of the many coefficients
that appear in the action with $R^3$ terms. But whatever the answer to that question, the question will
have to be repeated for yet higher order terms. The answer can also be sensitive to the
equation of state of other contributions to the energy momentum tensor. In the end
 a theory will either support the UV-IR coupled solutions or it will not; fine tuning is not necessary.

\begin{figure}
\centering
\includegraphics{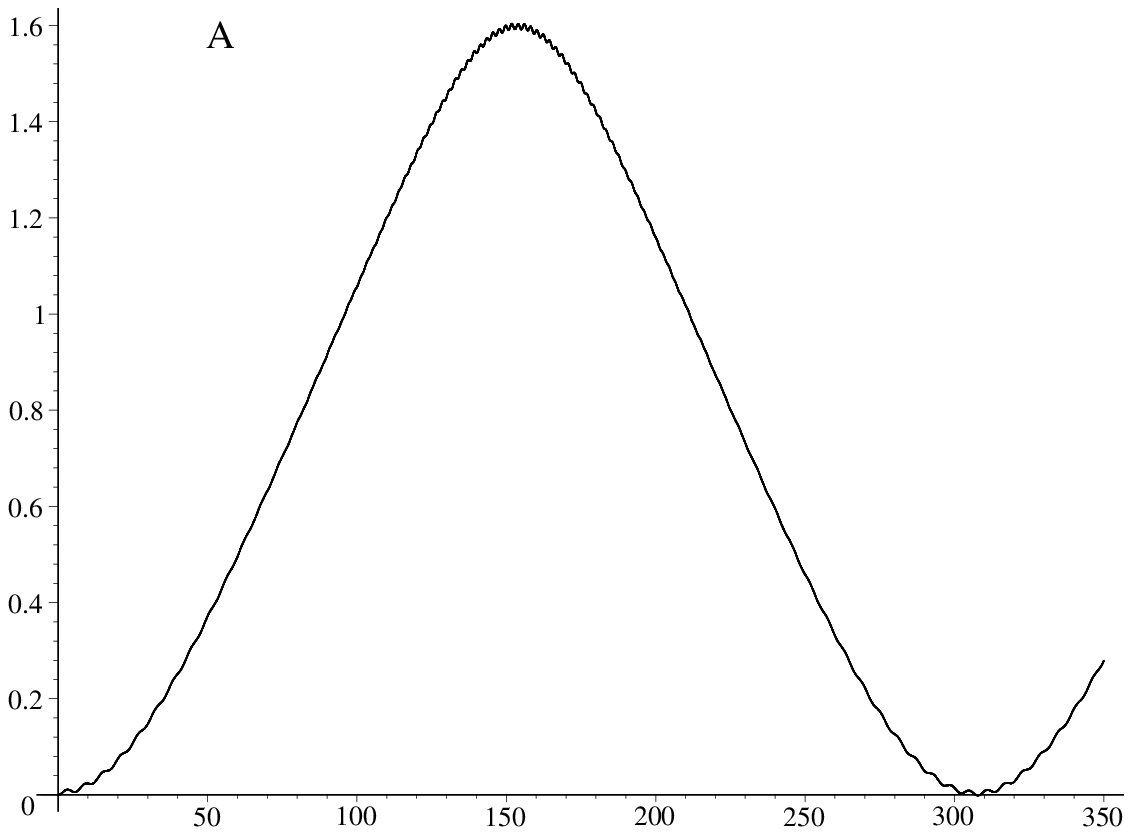}\vspace{10mm}
\includegraphics{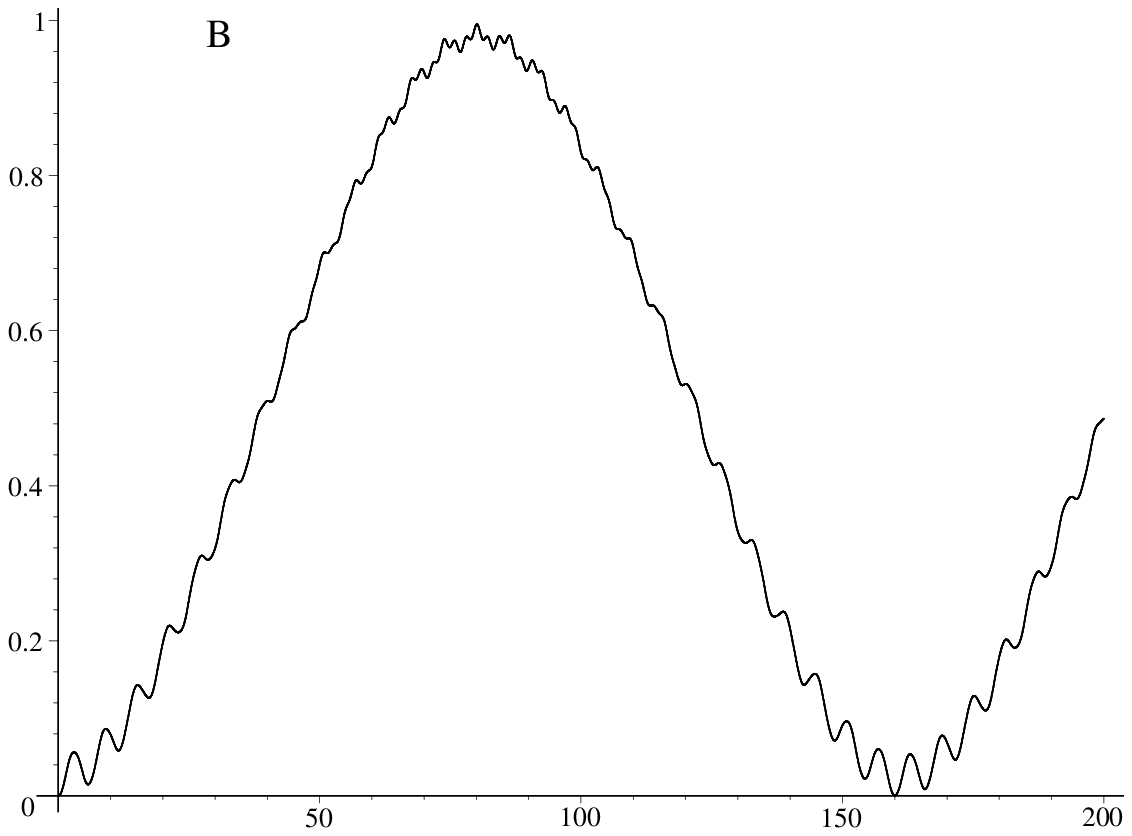}
\caption{Solutions to equation (\ref{ee}) with coefficients $d_1=.1$, $d_2=.2$, $d_3=.2$, 
$d_4=0$ and $\eta\varepsilon^2=10^{-3}/3$. The initial conditions generating A and B have vanishing derivatives
except for $A''(0)=0.005$ and 0.028 respectively. 
Solutions exist for a range of the initial conditions, including the case where all derivatives
in the initial conditions vanish.}
\end{figure}
Two examples of solutions to (\ref{ee}) with $\Lambda>0$ are displayed in Fig.~(2),
with the only difference being in the size of the initial conditions. 
For the larger initial derivatives, Fig.~(2B) shows that
the rapid oscillations can become irregular. This is presumably related to the existence of multiple
 rapidly oscillating modes associated with the linear terms, unlike
the case in (\ref{bb}). 

We can at least address the meaning of the condition on the linear order terms at arbitrarily
high order. The condition that the associated polynomial in the frequency has only real roots
is just the requirement that there are no exponentially evolving modes
among small fluctuations that retain the form of the metric in (\ref{dd}). 
When extended to all orders in the
derivative expansion of some underlying theory, this is the statement that the underlying 
theory is stable within this family of metrics.

For a theory that supports the UV-IR coupled solutions with $\Lambda>0$, there will
also be the de Sitter solution $A(\xi)\propto \xi$. But in this case the de Sitter solution
 is unstable.
Numerical analysis shows that the fast oscillations will tend to grow from any 
perturbation of the various derivatives of $A(\xi)$ to approach the
order $\varepsilon$ amplitude, and eventually the large scale expansion will turn into
a large scale oscillation. In other words, unlike the situation in (\ref{gg}), the de Sitter solution is not the 
generic solution. Of course in a different theory it may be that the de Sitter solution is generic,
rather than the UV-IR coupled ones. For example in (\ref{ee}) the de Sitter solution
is generic for $d_4$ sufficiently large and positive.

There is the related question of a possible decay \cite{c,d} of the fast oscillations to energetic particles,
occurring through the gravitational coupling. We see that this would involve a competition
between the decay and the tendency for the oscillations to grow. And since the metric is conformally
flat the decay would require conformal symmetry breaking, for example in the form 
of a particle mass, implying suppressed couplings of the oscillations to matter.\footnote{
In higher
dimensions the oscillating metric need not be conformally flat, resulting
in larger couplings \cite{d}.}

A decay to a de Sitter state would also assume that that is the preferred state.
This is by no means obvious. Given the role of nonlinearities in the equations, it seems more
likely that a more generic state is preferred where the various derivatives of $A(\xi)$ are all
of the same order. Thus there may well be some kind of decay to a ground state, 
but that ground
state could be a particular UV-IR coupled solution.

All of this says nothing about the problem of why the cosmological constant is so small.
This question becomes more interesting in some context in which the cosmological
constant is a derived quantity. This may occur in higher dimensional theories \cite{a},
and in particular in a singularity-free warped Kaluza-Klein picture \cite{b}
 where the effective 4D cosmological constant
is a function of the amount of warping.
In fact there are solutions \cite{c} where time dependent oscillations coexist with the warping of a fifth
dimension; and in this case the equations of the $R+R^2$ action yield $\dot{\phi}^2>0$. 
But solutions having an exponential
expansion as well were not found. This is understandable in light of the
present results which indicate that the generic solutions are oscillatory
on large time scales as well as short time scales. To actually demonstrate UV-IR coupled solutions within
the warped Kaluza-Klein framework would require solving nontrivially coupled high-order differential
equations.

Nevertheless, UV-IR coupled solutions may well exist in such a framework.
If such a solution is preferred for a given effective 4D cosmological
constant, then the oscillation and its coupling to normal matter may provide
a mechanism for the decay of the effective 4D cosmological
constant. Note that the metric would have to adjust on both large and small
time scales, since the relaxation would proceed through a series of UV-IR coupled solutions as
$\varepsilon$ decreased.

\section*{Acknowledgments}
I am grateful to H. Collins
for our collaborative work listed in the references. 
This research was supported in part by the Natural Sciences 
and Engineering Research Council of Canada.

\end{document}